\documentstyle[pra,aps,psfig,epsfig]{revtex}
\begin{document}
\draft

\title{Baryogenesis {\it vs.} proton stability in theories with extra dimensions}

\author{Antonio Masiero\thanks{E-mail address: masiero@sissa.it}, Marco Peloso\thanks{E-mail address: peloso@sissa.it}, Lorenzo Sorbo\thanks{E-mail address: sorbo@sissa.it}, and Rula Tabbash\thanks{E-mail address: rula@sissa.it}}

\address{Scuola Internazionale Superiore di Studi
Avanzati (S.I.S.S.A.),\\ Via Beirut $2-4$, $I-34014$ Trieste, Italy \\ and \\
Istituto Nazionale di Fisica Nucleare, Sez. di Trieste \\ Via Valerio $2$,
$I-34127$ Trieste, Italy}

\date{March 31, 2000}

\maketitle

\begin{abstract}
We propose a new scenario of baryogenesis in the context of theories with 
large extra dimensions. The baryon number is almost conserved at zero
temperature by means of a localization mechanism recently analyzed by
Arkani-Hamed and Schmaltz: leptons and quarks are located at two slightly 
displaced positions in the extra space, and this naturally suppresses the 
interactions which ``convert'' the latter in the former. We show that this is
expected to be no longer true when finite temperature effects are taken into
account. The whole scenario is first presented in its generality, without
referring to the bulk geometry or to the specific mechanism which may generate
the baryon asymmetry. As an example, we then focus on a baryogenesis model
reminiscent of GUT baryogenesis. The Sakharov out of equilibrium condition is 
satisfied by assuming nonthermal production of the bosons that induce baryon 
number violation. 
\end{abstract}


Preprint : SISSA 31/2000/EP

\def \dbar{\partial \hskip -7.0pt/\hskip +3.0pt}
\def \lta {\mathrel{\vcenter{\hbox{$<$}\nointerlineskip\hbox{$\sim$}}}}
\def \gta {\mathrel{\vcenter{\hbox{$>$}\nointerlineskip\hbox{$\sim$}}}}

\section{Introduction}

Despite the great success of Quantum Field Theory, a consistent scenario  where
gravity is also included still lacks. The most promising framework  that could
help in this task is string theory, whose consistency requires additional
dimensions beyond the standard $3+1\,$. This extra space is usually assumed to
be compact, with a small compactification radius of order $M_p^{-\,1}\,$.
However, it has been observed in ref. \cite{add2} that, having no test of
gravity below the millimeter scale, we do not really need such a tiny
compactification radius, provided the extra dimensions are accessible only to
gravitational interactions. The Standard Model degrees of freedom must indeed
be localized on a $3$ dimensional wall whose inverse thickness does not exceed
the scale of energy, of order $\mbox{TeV}\,$, we currently probe in accelerator
experiments.

The choice of such large compactification scale has the main goal of solving
(or at least of weakening) the hierarchy problem. Denoting by $V_n$ the volume
of the compact space -- assumed in \cite{add2} to have a trivial metric -- and
by $M$ the fundamental scale of gravity, the observed Planck mass is obtained
by the relation $M_p^2 = M^{n+2} V_n \,$. Under the condition $n \geq 2\;$, $M$
can be safely assumed to be very close to the electroweak scale, without 
conflicting with either cosmological, astrophysical, or laboratory bounds. 

Although considerably improving the standard situation, the above scenario
retains however some degree of fine tuning, connected to the largeness of the
quantity $V_n \cdot M^n\,$. A better result in this regard is provided by the
more recent work \cite{rs1}, where, due to the presence of cosmological
constants in the bulk and on two walls, the metric is nonfactorizable with an
exponential scaling in the extra space. This fact allows the achievement of
both a (phenomenologically) acceptable theory with just one extra dimension
(that could be even infinite in extension \cite{rs2}) and a more satisfactory
solution to the hierarchy problem.

There are some other aspects common to all of these theories besides the ones
discussed above. In particular, both proton stability and baryogenesis may be
problematic in models with very low fundamental masses.

For what concerns proton stability in Grand Unified Theories, the standard way
to achieve it is to increase the mass of the additional bosons up to about
$10^{15} - 10^{16}\:$GeV $\,$. In the framework of theories with 
extra--dimensions, an interesting mechanism has been suggested in ref. 
\cite{ahs}.  \footnote{See also \cite{aahdd,st,bd} for alternative suggestions.} 
In this paper, a dynamical mechanism for the localization of fermions on the
wall \cite{rs} is adopted: leptons and quarks are however localized at two 
slightly displaced positions in the extra space, and this naturally 
suppresses the interactions which ``convert'' the latter in the former.

However, the observed baryon asymmetry requires baryon number ($B$) violating
interactions to have been effective in the first stages of the evolution of the
Universe. In this paper we thus wonder how this last requirement can be
satisfied in a theory which adopts the idea of \cite{ahs}, to ensure proton
stability {\it now} and baryon production {\it in the past}. Our proposal is
that thermal corrections, which are naturally relevant at early times, may
modify the localization of quarks and leptons so to weaken the mechanism that
suppresses the $B$ violating interactions. \footnote{There exist other
proposals for baryogenesis in these  theories \cite{bd,dg,p}: in the work
\cite{bd}, after considering several bounds on baryogenesis with large extra
dimensions, a mechanism based on nonrenormalizable operators is
proposed; in ref. \cite{dg} baryon number is violated by  ``evaporation'' of
brane bubbles that carry a net baryonic charge into the  bulk, and the
matter--antimatter asymmetry can be due to a primordial  collision of our brane
with another one, that carried away the missing antimatter; in ref. \cite{p}
baryogenesis is obtained via leptogenesis, the  latter being due to the
existence of sterile neutrinos in the bulk.}

The plan of the work is the following. In section $2$ we review the mechanism
used in ref. \cite{ahs} to localize chiral fermions on a domain wall and to
suppress the rate of baryon number violating interactions. In our work,
however, we also take into account the finite thickness of the wall: this
enforces on the parameters of the model some bounds which are stronger than the
ones reported in ref. \cite{ahs}. In section $3$ we estimate the thermal
corrections to the parameters of the theory, and in particular to the function
that measures the suppression of the B violating interactions. Being our model
nonrenormalizable, a perturbative treatment can be meaningful only at low
energies. Anyhow, it is conceivable that the results we present in section $3$
can be a hint for the behavior of the theory at higher temperatures. The issue
of baryogenesis is faced in section $4$, where we consider a very simple
example reminiscent of GUT baryogenesis. In this mechanism, the baryon
asymmetry is achieved through the dacay of a boson, whose interactions violate
baryon number. In order for the model to work, the boson must be out of
equilibrium before decaying, and this is not obvious in theories with low
fundamental masses. In those theories, the Hubble parameter is indeed very low
at energies below the fundamental scale of gravity, that sets the natural
cutoff of the theory. We will thus consider a model where the bosons
responsible for the baryon asymmetry are produced nonthermally (for instance,
at the end of inflation) fulfilling thus naturally the out of equilibrium
requirement. After considering other bounds, such as the stability of the kink
under thermal corrections, we finally calculate the baryon asymmetry in a $B -
L$ conserving scheme. In the conclusions we discuss our results and their
possible future extensions.

\section{Localization}

A simple mechanism for localizing fermions on a wall has been recently 
revisited in ref. \cite{ahs}.

In this paper the idea is illustrated in the easiest case where only one extra
dimension is added to the usual four. The main ingredient that is needed is a
scalar field $\phi$ which couples to the fermionic field $\psi$ through the
full five dimensional Yukawa interaction $g\,\phi\,{\bar \psi}\,\psi$ and whose
expectation value $\langle \phi \rangle$ varies along the extra dimension, but
it is constant on our four-dimensional world. \footnote{In this way the VEV
$\,\langle \phi \rangle$ breaks the full translational invariance, as it is
needed to have a preferred direction orthogonal to the wall.}

It is possible to show \cite{rs} that in this case the fermionic field
localizes where its total mass $m=m_0 + g \, \langle \phi \rangle\,$ ($m_0$ is
the bare fermionic mass in the five dimensional theory) vanishes, i.e. on a
wall with three spatial dimensions characterized by a particular position $x_5$
in the transverse direction.

For definiteness, we consider the theory described by the lagrangian
\begin{eqnarray} \label{lagr}
{\cal L}_{\phi \psi} &=& {\bar \psi} \left( i \, \dbar_5 +
\frac{1}{{\widetilde M}_0^{1/2}} \, \phi \left( y \right) + m_0 \right)
\psi \nonumber\\
{\cal L}_\phi &=& \frac{1}{2} \partial_\mu \phi \, \partial^\mu \phi - \left(
-\,\mu_0^2 \, \phi^2 + \lambda_0 \, \phi^4 \right)\;\,,
\end{eqnarray}
where $y \equiv x_5$ is the fifth coordinate, the fields and the parameters 
have the following mass dimensions
\begin{equation}
\left[ \phi \right] = 3/2 \;,\; \left[ \psi \right] = 2 \;,\;
\left[ m_0 \right] = \left[ \mu_0 \right] = \left[ {\widetilde M}_0 \right]
= 1 \;,\; \left[ \lambda_0 \right] = -\,1 \;\;,
\end{equation}
and where the suffix $0$ indicates the value of the parameters at zero
temperature. 

As we said, the localization position of the fermions depends on the vacuum
configuration of the field $\phi\,$. If we consider the kink solution
\begin{equation}  \label{kink}
\phi = \frac{\mu_0}{\sqrt{2 \, \lambda_0}} \, \tanh \left( \mu_0 \, y
\right) \;\;,
\end{equation}
and we approximate it with a straight line interpolating between the two
vacua (see figure $1$)
\begin{eqnarray} \label{approx}
\cases{\phi \left( y \right) \simeq \frac{\mu_0^2}{\sqrt{2 \, \lambda_0}} \,
y \;\;,\;\; \vert y \vert < \frac{1}{\mu_0}  & \cr
\phi \left( y \right) \simeq \pm \frac{\mu_0^2}{\sqrt{2 \, \lambda_0}}
\;\;,\;\; \vert y \vert > \frac{1}{\mu_0} \;\;,&}
\end{eqnarray}

\begin{figure}[h]
\caption{}
\hspace*{4.2cm}
\epsfbox{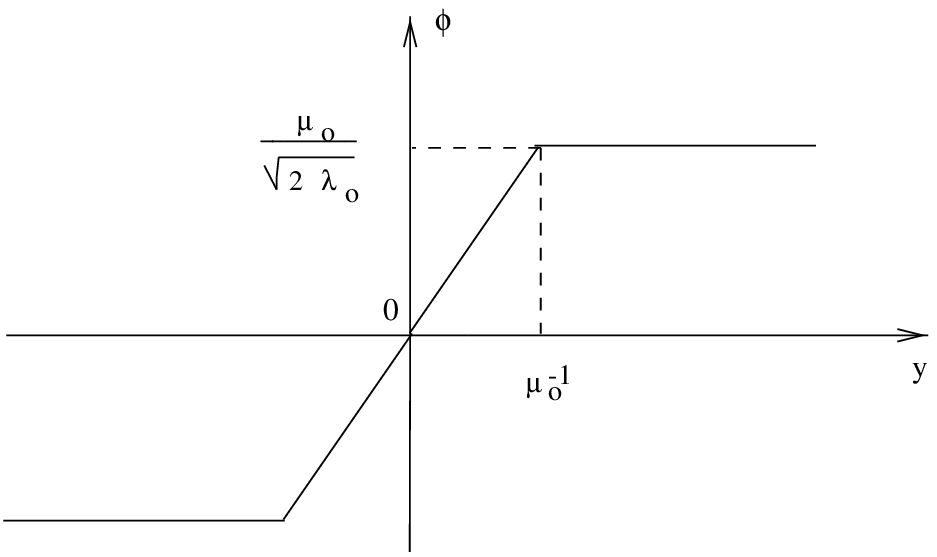}
\end{figure}
we see that the localization can occur only if
\begin{equation} \label{lcon}
m_0 < \frac{\mu_0}{\sqrt{2\,\lambda_0\,{\widetilde M}_0}} \;\;,
\end{equation}
since otherwise the total fermion mass
\begin{equation}
m_{\mathrm{tot}} = \frac{1}{{\widetilde M}_0^{1/2}} \, \phi \left( y \right)
+ m_0
\end{equation}
never vanishes.

It can be shown that, from the four dimensional point of view, a left handed
chiral massless fermionic field results from the localization mechanism, if the
above configuration (\ref{kink}) is assumed for the scalar $\phi\,$. The right
handed part remains instead delocalized in the whole space. This is not a
problem since it is customary to limit the Standard Model and the MSSM
fermionic content only to left handed fields. \footnote{Concerning the
cancellation of anomalies on the wall and recovering the Standard Model
running of the coupling constants, see \cite{rds}.}  The right handed fields can
also be localized  if a kink--antikink solution is assumed for the scalar
$\phi\,$. As a result, the left fields continue to be localized on the kink,
while the right ones are confined to the antikink. If the kink and the antikink
are sufficiently far apart, the left handed and right handed fermions however
do not interact and again the model reproducing our four dimensional world must
be built by fermions of a defined chirality. The fermionic content of the full
dimensional theory is in this case doubled with respect to the usual one, and
observers on one of the two walls will refer to the other as to a ``mirror
world''. The presence of this kink--antikink configuration may be required by
stability consideration if thermal effects are considered, as it is the case in
the next sections. However most of the physics in one brane is not affected by
the presence of the mirror one, and in the most of the present work we will
concentrate on a single wall as if only the kink (\ref{kink}) configuration was
present.

In order to give mass to the fermions, some other scalar field acting as a
Higgs in the four dimensional theory must be considered. As it is shown in ref.
\cite{ahs}, the mechanism described above could give an explanation to the
hierarchy among the Yukawa couplings responsible for the fermionic mass matrix.
If indeed one chooses different five dimensional bare masses for the different
fermionic fields, the latter are localized at different positions in the fifth
direction. As a consequence, the wave functions of different fermions do only
partially overlap, and increasing the difference between the five dimensional
bare masses of two fermions results in suppressing their mutual interactions.

The same idea can be adopted to guarantee proton stability. Let us give,
respectively, leptons and baryons the ``masses'' $$ \left( m_0 \right)_l = 0
\;\;,\;\; \left( m_0 \right)_b = m_0\;,$$ which correspond to the localizations
\footnote{The last inequality in the next expression comes from (\ref{lcon}). We assume quarks of different generations to be located in the same $y$ position in order to avoid dangerous FCNC mediated by the Kaluza-Klein modes of the gluons \cite{dpq}.} 
$$ y_l = 0 \;\;,\;\;
y_b = \frac{m_0 \, \sqrt{2\,\lambda_0\,{\widetilde M}_0}}{\mu_0^2} <
\frac{1}{\mu_0} $$ The shape of the fermion wave functions along the fifth dimension can be cast in an explicit and simple form if we
consider the limit $y_b \ll 1/\mu_0$, in which the effect of the
plateau for $y>1/\mu_0$ can be neglected: \footnote{This is also the limit in
which the approximation (\ref{approx}) is valid.}
\begin{eqnarray}
f_l \left( y \right) &=& \left( \frac{\mu_0^2}{\sqrt{2\,\lambda_0\,
{\widetilde M}_0} \pi} \right)^{1/4} \: \exp \left \{ {-\,\frac{\mu_0^2\,y^2}
{2\,\sqrt{2\,\lambda_0\,{\widetilde M}_0}}}\right \} \nonumber\\
f_b \left( y \right) &=& \left(\frac{\mu_0^2}{\sqrt{2\,\lambda_0\,
{\widetilde M}_0} \pi} \right)^{1/4} \: 
\exp \left \{{-\,\frac{\mu_0^2\,\left(y-y_b\right)^2}{2\, 
\sqrt{2\,\lambda_0\,{\widetilde M}_0}}}\right \} \;\;.
\end{eqnarray}

We assume the Standard Model to be embedded in some theory which, in general,
contains some additional bosons $X$ whose interactions violate baryon number
conservation. If it is the case, the four fermion interaction $qq
\longleftrightarrow ql$ can be effectively described by
\begin{equation} \label{sca}
\int d^4 x \, d y \:\frac{q\,q\,q\,l}{\Lambda \, m_X^2}\;\;,
\end{equation}
where $m_X$ is the mass of the intermediate boson $X$ and $\Lambda$ is a
parameter of mass dimension one related to the five-dimensional coupling of the
X-particle to quarks and leptons.

This scattering is thus suppressed by
\begin{eqnarray} \label{supp}
I &=& \frac{1}{\Lambda \,m_X^2} \int dy \frac{\mu_0^2}{\pi \sqrt{2\,\lambda_0\,
{\widetilde M}_0} }\: \exp \left \{ -\,\frac{\mu_0^2 /2}{\sqrt{2\,\lambda_0\,
{\widetilde M}_0}} \, \left[ y^2 + 3 \, \left( y - y_b \right)^2 \right]\right
\} = \nonumber \\
\dots &=& \frac{\mu_0}{\Lambda \, m_X^2 \, \sqrt{2\,\pi} \, \left(
2\,\lambda_0\,{\widetilde M}_0 \right)^{1/4}} \: \exp \left \{{-\frac{3\,
\left( 2\,\lambda_0\,{\widetilde M}_0 \right)^{1/2}}{8}\,\frac{m_0^2}
{\mu_0^2}}\right \}\;\;.
\end{eqnarray}

Current proton stability requires $I \lta \left(10^{16} \,\mbox{GeV} \right)^{-\,2}\;,$
that is 
\begin{equation} \label{con2}
\frac{m_0}{\mu_0} \gta \frac{\sqrt{200-6\,\mbox{Log}_{10}\left(\frac{\Lambda\,
m_X^2}{\mu_0}\Big / \mbox{GeV}^2\right )}}{\left( 2\,\lambda_0\,
{\widetilde M}_0 \right)^{1/4}}\;\;.
\end{equation}
The numerator in the last equation is quite insensitive to the mass scales of
the model, and -- due to the logarithmic mild dependence -- can be safely
assumed to be of order $10\,$. For definiteness, we will thus fix it at the
value of $10$ in the rest of our work. 

Conditions (\ref{lcon}) and (\ref{con2}) give altogether
\begin{equation}
\frac{10\,\mu_0}{\left( 2\,\lambda_0\,{\widetilde M}_0 \right)^{1/4}}
\lta m_0 \lta \frac{\mu_0}{\left( 2\,\lambda_0\,{\widetilde M}_0
\right)^{1/2}} \;\;,
\end{equation}
that we can rewrite
\begin{eqnarray} \label{limits}
\cases{2\, \lambda_0 \, {\widetilde M}_0 \lta 10^{-4} & \cr
\frac{m_0}{\mu_0} \gta 10^2 \;\;.}
\end{eqnarray}

The last limit in eqs. (\ref{limits}) is stronger than the one given in ref.
\cite{ahs} where proton stability is satisfied if the ratio of the massive
scales of the model is of order $10\,$. However, in ref. \cite{ahs} the field
$\phi$ simply scales linearly as a function of $y\,$, while we expect that
whenever a specific model is assumed, conditions analogous to our (\ref{lcon})
and (\ref{limits}) should be imposed.

\section{Thermal correction to the coefficients}
Once the localization mechanism is incorporated in a low energy effective
theory -- as the system (\ref{lagr}) may be considered --, one can legitimately
ask if thermal effects could play any significant role. In the present work we
are mainly interested in any possible change in the argument of the exponential
in eq. (\ref{supp}), that will be the most relevant for the purpose of
baryogenesis. For this reason, we introduce the dimensionless quantity
\begin{equation}\label{defa}
a(T)=\frac {m(T)^2}{\mu(T)^2}\sqrt{2\,\lambda (T) {\widetilde M}(T)}\;\;\;.
\end{equation}

From eqs.(\ref{con2}) and (\ref{limits}), we can set $a(0)\gta 100$ at zero
temperature. Thermal effects will modify this value. There are however some
obstacles that one meets in evaluating the finite temperature result. Apart
from some technical difficulties arising from the fact that the scalar
background is not constant, the main problem is that nonperturbative effects
may play a very relevant role at high temperature. As it is customary in
theories with extra dimensions, the model (\ref{lagr}) is nonrenormalizable and
one expects that there is a cut-off (generally related to the fundamental scale
of gravity) above which it stops holding. Our considerations will thus be valid
only for low temperature effects, and may only be assumed as a rough indication
for what can happen at higher temperature.

Being aware of these problems, by looking  at the dominant
finite-temperature one-loop effects, we estimate the first corrections to
the relevant parameters to be 
\begin{eqnarray} \label{thermal}
\cases{\lambda \left( T \right) = \lambda_0 + c_\lambda \, \frac{T}{{\widetilde
M}_0^2} & \cr {\widetilde M} \left( T \right) = {\widetilde M}_0 +
c_{{\widetilde M}} \, T & \cr m \left( T \right) = m_0 + c_m \,
\frac{T^2}{{\widetilde M}_0} & \cr
\mu^2 \left( T \right) = \mu_0^2 + c_{\mu} \, \frac{T^3}{{\widetilde M}_0}
\;\;,}
\end{eqnarray}
where the $c$'s are dimensionless coefficients whose values are related to the
exact particle content of the theory.

In writing the above equations, the first of conditions (\ref{limits}) has also
been taken into account. For example, both a scalar and a fermionic loop
contribute to the thermal correction to the parameter $\lambda_0\,$. While the
contribution from the former is of order $\lambda_0^2 \, T\,$, the one of the
latter is of order $T / {{\widetilde M}_0}^2$ and thus dominates.
\footnote{Notice also that in our model loops with internal leptons
dominate over loops with internal quarks, since the former have vanishing five
dimensional bare mass and thus are not Boltzmann suppressed. However, although
this choice is the simplest one, one may equally consider the most general case
where all the fermions have a nonvanishing five dimensional mass.}

Substituting eqs. (\ref{thermal}) into eq. (\ref{defa}), we get, in the limit
of low temperature,

\begin{equation}
a \left( T \right) \simeq a\left(0\right) \cdot \left[ 1 +
\frac{T}{{\widetilde M}_0} \: \left(
\frac{c_\lambda}{2\,\lambda_0\,{\widetilde M}_0} + \frac{c_{\widetilde
M}}{2} + \frac{2\,c_m\,T}{m_0} - \frac{c_{\mu}\,T^2}{\mu_0^2}
\right) \right] \;\;.
\end{equation}

From the smallness of the quantity $\lambda_0 \, {\widetilde M}_0\;$
[see cond. (\ref{limits})] we can safely assume (apart from high
hierarchy between the $c$'s coefficients that we do not expect to
hold) that the dominant contribution in the above expression comes
from the term proportional to $c_\lambda\,$.

We thus simply have
\begin{equation} \label{acorr}
a \left( T \right) \simeq a\left(0\right) \left( 1 + c_{\lambda} \,
\frac{T}{2\,\lambda_0\,{\widetilde M}_0^2} \right) \:\,.
\end{equation}

We notice that the parameter $c_\lambda$, being related to the thermal
corrections to the $\phi^4$ coefficient due to a fermion loop, is expected to
be {\it negative} \cite{ddgr}: the first thermal effect is to decrease the
value of the parameter $a(T)$, making hence the baryon number violating
reactions more efficient at finite rather than at zero temperature.

There is another effect which may be very crucial at finite temperature, linked
to the stability of the $Z_2$ symmetry. When a temperature is turned on, we
generally expect the formation of a fermion--antifermion condensate $\langle
{\bar \psi} \: \psi \rangle \neq 0\,$. If it is the case, the Yukawa coupling
$\phi\,{\bar \psi}\,\psi$ in the lagrangian (\ref{lagr}) renders one of the two
vacua unstable. While this leads to an instantaneous decay of the kink
configuration, a kink--antikink system could have a sufficiently long lifetime
provided the two objects are enough far apart.

\section{Baryogenesis}

We saw in the previous section that thermal effects may increase the rate of
baryon number violating interactions of our system. This is very welcome, since
a theory which never violates baryon number cannot lead to baryogenesis and
thus cannot reproduce the observed Universe. Anyhow baryon number violation is
only one of the ingredients for baryogenesis, and the aim of this section is to
investigate how the above mechanism can be embedded in a more general context.

A particular scheme which may be adopted is baryogenesis through the decay of
massive bosons $X$. \footnote{We may think of these bosons as the intermediate
particles which mediate the four fermion interaction described by the term
(\ref{sca}).} This scheme closely resembles GUT baryogenesis, but there are
some important peculiarities due to the different scales of energy involved. In
GUT baryogenesis the massive boson $X\;$, coupled to matter by the interaction
$g\,X\,\psi\,{\bar \psi}\,$, has the decay rate
\begin{equation}
\Gamma \simeq \alpha \, m_x \;\;,\;\; \alpha = \frac{g^2}{4\,\pi} \;\;. 
\end{equation}

An important condition is that the $X$ boson decays when the temperature of the Universe is below its mass (out of equilibrium decay), in order to avoid thermal regeneration. From the standard equation for the expansion of the Universe,
\begin{equation}
H \simeq g_*^{1/2}\,\frac{T^2}{M_p} \;\;
\end{equation}
(where $g_*$ is the number of relativistic degrees of freedom at the
temperature $T$), this condition rewrites
\begin{equation} \label{out}
m_X \gta g_*^{-1/2}\,\alpha \, M_p \;\;.
\end{equation}

If $X$ is a Higgs particle, $\alpha$ can be as low as $10^{-\,6}\,$. Even in
this case however the $X$ boson must be very massive. In principle this may be
problematic in the theories with extra dimensions we are interested in, which
have the main goal of having a very low fundamental scale. 

There are some possibilities to overcome this problem. One is related to a
possible deviation of the expansion of the Universe from the standard behavior.
The issue of the Friedmann law in models with large extra dimensions has been
indeed subject of intense debate in the recent past. In the work \cite{bdl} a
detailed analysis of the Einstein equations with one extra dimension, shows
that the expansion rate $H$ should (in absence of any energy in the bulk) be
proportional to the energy density $\rho$ on the brane. This behavior strongly
conflicts with the standard one $H \propto \rho\,^{1/2}\,$. In refs.
\cite{cgkt,cgs,sms} it was then shown that the standard expansion law could be
achieved, at least at low temperatures, by a suitable fine tuning of the vacuum
energies in the Universe. In particular, this is the case for the
Randall--Sundrum model \cite{rs1}, which offers one of the most satisfactory
solutions to the hierarchy problem. The solution proposed in these works
\cite{cgkt,cgs,sms} is however itself plagued by some other cosmological
problems. For example, in the RS model \cite{rs1} gravity turns out to be
repulsive on our brane. All the above problems are overcome when some mechanism
for the stabilization of the radion is taken into account, as the analyses
\cite{cgrt,kkop} show. In particular, in ref. \cite{cgrt}, the case of the
model \cite{rs1} is examined, and both the standard Friedmann law and the
``correct'' sign for the Newton constant are obtained. The analysis of ref.
\cite{cgrt} is however performed by computing only the first-order term of the
expansion of the square of the Hubble parameter $H^2$ as a power series of the
energy density $\rho$. Terms of order $\rho^2$ could become relevant at
temperatures above $1\, \mbox{TeV}$ or so \cite{cgrt}. This may result in an
accelerated expansion of the Universe at high temperatures, and the out of
equilibrium condition for the $X$ bosons could be consequently considerabily
favoured.

However, both the facts that we do not know the exact behavior of the expansion
rate at high temperatures in the model \cite{rs1}, and that one may be
interested in embedding our baryogenesis mechanism in some other cosmological
scenario, lead us to discuss alternative solutions for the out of equilibrium
problem. One very natural possibility is to create the $X$ particles non
thermally and to require the temperature of the Universe to be always smaller
than their mass $m_X\,$. In this way, one kinematically forbids regeneration of
the $X$ particles after their decay. In addition, although interactions among
these bosons can bring them to thermal equilibrium, chemical equilibrium cannot
be achieved.

Nonthermal creation of matter has raised a considerable interest in the last years. In particular, it has been shown that this production can be very efficient during the period of coherent oscillations of the inflaton field after inflation \cite{tb,kls,stb}. The efficiency of this mechanism has also been exploited in the work \cite{lkr} to revive GUT baryogenesis in the context of  standard four dimensional theories. Here, we will not go into the details of the processes that could have lead to the production of the $X$ bosons. Rather, we will simply assume that, after inflation, their number density is $n_X\,$. To simplify our computations, we will also suppose that their energy density dominates over the thermal bath produced by the perturbative decay of the inflaton field.

Just for definiteness, let us consider a very simple model where there are two species of $X$ boson which can decay into quarks and leptons, according to the four dimensional effective interactions
\begin{equation} \label{deca}
g\,X\,{\bar q}\,{\bar q} \;\;\;,\;\;\; g\,e^{-\,a/4}\,X\, l\,q \;\;,
\end{equation} 
where (remember the suppression given by the different localization of quarks
and leptons) the quantity $a$ is defined in eq. (\ref{defa}).
Again for definiteness we will consider the minimal model where no extra fermionic degrees of freedom are added to the ones present in the Standard Model. Moreover we will assume $B - L$ to be conserved, even though the extension to a more general scheme can be easily performed.

The decay of the $X$ bosons will reheat the Universe to a temperature that can be evaluated to be
\begin{equation}
T_{\mathrm {rh}}\simeq
\left(\frac{30}{\pi^2}\,\frac{m_X\,n_X}{g_*}\right)^{1/4}\,\,.
\end{equation}

Since we do not want the $X$ particles to be thermally regenerated after their decay, we require $T_{\mathrm {rh}}\lta m_X$, that can be rewritten as an upper bound on $n_X$
\begin{equation}
n_X\lta 30\,\left(\frac{g_*}{100}\right)\,m_X^3\,\,.
\end{equation}

Another limit comes from the necessity to forbid the $B$ violating four fermion interaction (\ref{sca}) to erase the $B$ asymmetry that has been just created by the decay of the $X$ bosons. We thus require the interaction (\ref{sca}) to be out equilibrium at temperatures lower than $T_{\mathrm {rh}}$. From eq. (\ref{supp}) we see that we can parametrize the four fermion interaction with a coupling $g^2\,e^{-3\,a/8}/m_X^2$. Hence, the out of equilibrium condition reads
\begin{equation}\label{outfour}
g^4\,e^{-3\,a/4}\lta g_*\, \frac{m_X}{M_p}\,\left(\frac{m_X}{T_{\mathrm {rh}}}\right)^3\;\;.
\end{equation}

One more upper bound on the reheating temperature comes from the out of equilibrium condition for the sphalerons.  
This requirement is necessary only if one chooses the theory to be $B - L$
invariant, while it does not hold for $B - L$ violating schemes. We can approximately consider the sphalerons to be in thermal equilibrium at temperatures above the electroweak scale. Thus, if $B - L$ is a conserved quantity, we will require the reheat temperature to be smaller than about $100\, {\mathrm {GeV}}$. 

If one neglects the presence of the thermal bath prior to the decay of the $X$
bosons, the very first decays will be only into couples of quarks, since the 
channel into one quark and one lepton is strongly suppressed by the
$e^{-a\left(T=0\right)}$ factor due to the fact that the kink is not modified
by any thermal correction. However, the decay process is not an instantaneous
event. It is shown in ref. \cite{ckr} that the particles produced in the very
first decays are generally expected to thermalize very rapidly, so to create a
thermal bath even when most of the energy density is still stored in the
decaying particles. \footnote{As shown in ref. \cite{ckr}, what is called the
reheating temperature is indeed the temperature of the thermal bath when it starts to dominate. After the first decays, the temperature of the light degrees
of freedom can be even much higher than $T_{\mathrm {rh}}$.} The temperature
of this bath can even be considerably higher than the final reheating
themperature. The presence of the heat bath modifies in turn the shape of the
kink, as shown in the previuos section, and we can naturally expect that this
modification enhances the $B$ violating interaction.

If the energy density of the Universe is dominated by the $X$ bosons before they decay, one has 
\begin{equation}\label{etabar}
\eta_B\simeq\,0.1 \,\left(N_X\, T_{\mathrm {rh}}/m_X\right)\,\,\langle r-{\bar r}\rangle\;\;,
\end{equation}
where $N_X$ is the number of degrees of freedom
associated to the $X$ particles and $\langle r-{\bar r}\rangle$ is the
difference between the rates of the decays $X\rightarrow q\, l$ and $\bar
X\rightarrow\bar q\, \bar l$.

We denote with $X_1$ and $X_2$ the two species of bosons whose interactions (\ref{deca}) lead to baryon number violation, and parametrize by $\epsilon$ the strength of CP-violation in these interactions. Considering that $e^{-2a}$ is always much smaller than one, we get \cite{nw}

\begin{equation}
\langle r-{\bar r}\rangle\sim 3\, g^2\, e^{-a/2}\,\epsilon\; {\mathrm
Im}\,\mbox{I}_{SS}\left(M_{X_1}/M_{X_2}\right)\,\:,
\end{equation}
where the function ${\mathrm Im}\,\mbox {I}_{SS} (\rho)=[\,\rho^2\;{\mathrm
Log}(1+1/\rho^2)-1\,]\,/\,(16\,\pi)$ can be estimated to be of order $10^{-3} -
10^{-2}$. It is also reasonable to assume $\epsilon \sim 10^{-2} - 1$. 

Collecting all the above estimates, and assuming $N_X$ to be of order $10$, we get 

\begin{equation}\label{etabaryon}
\eta_B\simeq \left(10^{-5}\,-\,10^{-2}\right)\, g^2 \, \frac{T_{\mathrm {rh}}}{m_X}\,e^{-a\left(T_{\mathrm {rh}}\right)/2}\;\;.
\end{equation}

From the requirement $T_{\mathrm{rh}}\lta m_X$ we get an upper limit on the baryon asymmetry
\begin{equation}\label{limit1}
\eta_B\lta \left(10^{-5}\,-\,10^{-2}\right)\,g^2\,e^{-a/2}\,\,,
\end{equation}
where the factor $a\left( T\right)$ has to be calculated for a value of $T$ of the order of the reheating temperature. 

We get a different limit on $\eta_B$ from the bound (\ref{outfour}): assuming $m_X\sim \mbox{TeV}$ and $g_*\sim 100$ indeed one obtains

\begin{equation}\label{limit2}
\eta_B\lta\left(10^{-6}\,-\,10^{-10}\right)\,g^{2/3}\,e^{-a/4}\,\,.
\end{equation}

Since the observed amount of baryon asymmetry is of order $10^{-10}$, even in the case of maximum efficiency of the process (that is, assuming maximal $CP$ violation and $g\sim 1$), we have that both bounds (\ref{limit1}) and (\ref{limit2}) imply that $a\left(T_{\mathrm {rh}} \right)$ has to be smaller than about $40$. 

Unfortunately, the temperature at which the condition $a \left( T \right) \lta 40\:$ occurs cannot be evaluated by means of the expansion of eq. (\ref{acorr}), that have been obtained under the assumption $\left | a\left ( T\right )-a\left ( 0 \right )\right | \ll a\left ( 0 \right )$. On the other hand, it is remarkable that our mechanism may work with a ratio $a\left ( T_d\right ) / a\left ( 0 \right )$ of order one. We thus expect that a successful baryogenesis may be realized for a range of the parameters of our theory which -- although not evaluable through a perturbative analysis -- should be quite wide and reasonable.

In scenarios with large extra dimensions and low scale gravity, the maximal
temperature reached by the Universe after inflation is strongly bounded from
above in order to avoid overproducing Kaluza-Klein graviton modes, which may
eventually contradict cosmological observations \cite{add}. For instance,
in models with two large extra dimensions the reheating temperature has to be
less than about $10\, \mbox{MeV}$. This value would be too low for our scenario
since $\eta_B$ is proportional to the ratio $T_{\mathrm {rh}}/m_X$, and hence
the observed amount of baryons would be reproduced at the price of an
unnaturally small value of $a\left( T_{\mathrm {rh}}\right)$. However, other
schemes with extra dimensions exist where the bounds on $T_{\mathrm {rh}}$ are
less severe. For example, in the proposals \cite{rs1,kmst} the mass of the
first graviton KK mode is expected to be of order TeV. The reheating
temperature can thus safely be taken to be of order $10-100\, \mbox{GeV}$. 

An alternative way to overcome the bound (\ref{out}) relies on the fact that, as observed in the work \cite{ckr}, the maximal temperature reached by the thermal bath during reheating can indeed be much higher than the final reheating temperature. In this case, even if $T_{\mathrm {rh}}$ is considerably lower than $m_X$, $X$ particles can be produced in a significant amount, and the out of equilibrium condition is easily achieved. However, the treatment of this mechanism is in our case somewhat different from the one given in ref. \cite{ckr}: due to the slowness of the expansion of the Universe, the $X$ bosons will decay before the freeze out of their production. The final baryon asymmetry cannot be estimated with the use of the formulae of \cite{ckr}, which are valid only if the decay of the $X$ particles occurs well after their freeze out.

There are of course several possible baryogenesis schemes alternative to the one just presented. A possible option which also requires a minimal extension to the Standard Model could be to achieve the baryon asymmetry directly through the $4$ fermions interactions $q + q \leftrightarrow q + l$ in the thermal primordial bath. The out of equilibrium condition may be provided by the change of the kink as the temperature of the bath decreases. \footnote{This condition may be easily achieved due of the exponential dependence of the rate of this process on the temperature, see eq. (\ref{supp}).} What may be problematic is the source of $CP$ violation which may lead the creation of the baryon asymmetry. A possibility in this regard may be provided by considering a second Higgs doublet, but the whole mechanism certainly deserves a deep analysis on its own.

\section{Conclusions}

The present work concerns the important issue of baryogenesis in theories with
large extra--dimensions. Since the observed proton stability requires to a very
high degree of accuracy baryon conservation at zero temperature, this task may
be problematic within the above theories, which have very low fundamental
scales.

Our proposal relies on the localization mechanism for fermions discussed in
ref. \cite{ahs}. While in this work the present proton stability is due to a
different localization (in the transverse direction) of leptons and quarks, we
believe that thermal corrections may activate early baryon violating
interactions.

In our work we first provide a general discussion of the above scheme, without
referring to any particular mechanism of baryogenesis. We find indeed that the
first thermal corrections are in the direction of increasing the rate of baryon
violations. 

We then consider a very specific example, where the matter--antimatter
asymmetry is achieved through the decay of a (relatively) heavy boson in a $B -
L$ conserving context. In this situation the Sakharov out of equilibrium condition 
can be obtained in the simplest way by considering nonthermal production 
of the bosons responsible for $B + L$ violation.

Several bounds apply to the whole mechanism. The most general ones concern the
localization procedure (we have found that the limits given in ref. \cite{ahs}
become more stringent once the thickness of the wall is considered) and its
stability against thermal corrections. In addition, there are some other
constraints which hold in the particular scheme of baryogenesis we adopted. 
The temperature of the heat bath right after the production of the baryon 
asymmetry cannot be too high, to avoid thermal regeneration 
of the bosons that induced baryogenesis. Moreover, this temperature has not 
to exceed the electroweak scale, in order not to activate the sphaleron 
transitions that would erase the $B + L$ asymmetry produced at some higher 
energy. Of course, this last bound can be easily overcome by considering some 
$B - L$ nonconserving process.

We have found that the observed baryonic asymmetry can be accomplished quite
naturally in our example, and we believe this should be the case in a more
general context as well. 

Possible extensions of the present work are related to the generality of the
scenario we discussed. Our idea indeed relies only on the localization
mechanism adopted in \cite{ahs} and not on the geometry of the bulk, nor on the
details of the interactions responsible for the baryon asymmetry.

Future works could thus proceed in two directions. Firstly, one could try to
embed the scheme here described in a more complete cosmological setting.
Secondly, some other baryogenesis mechanisms, for example {\it \`a la}
Affleck--Dine (which does not require very high energy scales), may be
explored.

\acknowledgments{
We thank J.\ March--Russell, G.\ Mussardo, H.P.\ Nilles, M.\ Pietroni, and A.\ Riotto for interesting and stimulating discussions. This work is partially supported by the MURST research project ``Astroparticle Physics''.}

\end{document}